\documentclass[twocolumn,showpacs,preprintnumbers,amsmath,amssymb]{revtex4}

\usepackage{graphicx}
\usepackage{dcolumn}
\usepackage{bm}

\begin{document}
\sloppy
\preprint{preprint}
\title {Anisotropic spin fluctuations in heavy-fermion superconductor CeCoIn$_5$:\\ In-NQR and Co-NMR studies}
\author{Y.~Kawasaki}
\altaffiliation[Present address: ]{Department of Physics, Faculty of Engineering, Tokushima University, Tokushima 770-8506, Japan}
\author{S.~Kawasaki}%
\author{M.~Yashima}%
\author{G.-q.~Zheng}%
\author{Y.~Kitaoka}%
\affiliation{%
Department of Physical Science, Graduate School of Engineering Science, Osaka University, Toyonaka, Osaka 560-8531, Japan}%

\author{H.~Shishido}
\author{R.~Settai}
\affiliation{%
Department of Physics, Graduate School of Science, Osaka University, Toyonaka, Osaka 560-0043, Japan}%

\author{Y.~Haga}
\affiliation{%
Advanced Science Research Center, Japan Atomic Energy Research Institute, Tokai, Ibaraki 319-1195, Japan}%

\author{Y.~\=Onuki}
\affiliation{%
Department of Physical Science, Graduate School of Engineering Science, Osaka University, Toyonaka, Osaka 560-8531, Japan}%
\affiliation{%
Advanced Science Research Center, Japan Atomic Energy Research Institute, Tokai, Ibaraki 319-1195, Japan}%


\date{\today}
\begin{abstract}
We report In-NQR and Co-NMR experiments of CeCoIn$_5$ that undergoes a superconducting transition with a record high $T_{\rm c}$ = 2.3 K to date among heavy-fermion superconductors.
At zero magnetic field, an anomalous temperature ($T$) dependence of nuclear spin-lattice relaxation rate $1/T_1$ of $^{115}$In is explained by the relation $1/T_1\propto T\cdot\chi_Q(T)^{3/4}$ based on the anisotropic spin-fluctuations model in case of the proximity to an antiferromagnetic (AFM) quantum critical point (QCP).
The novel behavior of $1/T_1\sim T^{1/4}$ over a wide $T$ range of $T_{\rm c} < T <  40$ K arises because the staggered susceptibility almost follows the Curie law $\chi_Q(T)\propto 1/(T+\theta)$ with $\theta= 0.6$ K and hence $1/T_1\propto T/(T+0.6)^{3/4}\sim T^{1/4}$ for $\theta < T$.
We highlight that the behavior $1/T_1\sim T^{1/4}$ is due to the proximity to the anisotropic AFM QCP relevant with its layered structure, and is not associated with  the AFM QCP for isotropic 3D systems.
We have also found that the AFM spin fluctuations in CeCoIn$_5$ are suppressed by small magnetic field so that $\theta=0.6$ K at $H$=0 increases to $\theta = 2.5$ K at $H$ = 1.1 T, reinforcing that CeCoIn$_5$ is closely located at the QCP.\end{abstract}
\maketitle

\section{Introduction}
Heavy-Fermion (HF) superconductivity has been a matter of interest with respect to an intimate interplay between magnetism and superconductivity since the discovery of HF superconductivity in CeCu$_2$Si$_2$ \cite{steglich79}.
From the extensive experimental and theoretical works on CeCu$_2$Si$_2$, it has been revealed that the unconventional superconducting (SC) phase takes place at the border of an antiferromagnetic (AFM) phase and even coexists with it when the superconductivity occurs very close to the AFM phase.
CeCu$_2$Si$_2$ was the only cerium-based HF superconductor at ambient pressure, before a new family of HF compounds CeIrIn$_5$ and CeCoIn$_5$ forming in two-dimensional (2D) tetragonal structures was discovered to show superconductivity below $T_{\rm c}$ = 0.4 K and 2.3 K, respectively \cite{petrovic01,petrovic01a}.
Remarkably, the value of $T_{\rm c} = 2.3$ K for the latter is a record of high $T_{\rm c}$ to date among previous examples.

Meanwhile, the HF antiferromagnet CeRhIn$_5$, which crystallizes in the same 
crystal structure, was discovered to show superconductivity below $T_{\rm c}$ = 2.2 K at pressures exceeding a critical pressure of 1.63 GPa \cite{hegger00}.
In CeMIn$_5$ (M = Ir, Co, and Rh), the unconventional superconductivity with line-node gap is suggested from the power-law temperature ($T$) dependence of specific heat \cite{petrovic01,petrovic01a,fisher02}, nuclear relaxation rate $1/T_1$ \cite{mito01,kohori01,curro01,zheng01}, penetration depth \cite{higemoto02,ormeno02,ozcan02}, thermal conductivity \cite{izawa01}, and so on.
Systematic investigation of this new family of HF superconductors allows us to unravel a relationship between the occurrence of unconventional superconductivity and possible spin fluctuations in the normal state.

In fact, a number of experiments point to non-Fermi-liquid (NFL) behavior above $T_{\rm c}$ in these compounds.
For instance, in CeCoIn$_5$, the physical quantities such as the specific heat divided by temperature $C/T$, the Pauli susceptibility $\chi$ and $1/T_1T$ reveal a significant increase upon cooling without entering any conventional Fermi-liquid regime where these quantities stay constant \cite{petrovic01,kohori01}.
The $T$ dependence of resistivity does not follow a conventional $T^2$ dependence but indicate approximately a $T$-linear dependence \cite{sidorov02}.
In the magnetic-field ($H$) induced normal state, $C/T$ continues to increase upon cooling down to 50 mK.
These NFL behaviors are considered to originate from magnetic fluctuations due to the proximity to an AFM quantum critical point (QCP)\@.
In CeCoIn$_5$, therefore, the unconventional superconductivity arises from the NFL normal state.

The AFM spin fluctuations in CeIrIn$_5$ and CeRhIn$_5$ have been investigated by inelastic neutron diffraction \cite{bao02} and nuclear quadrupole resonance (NQR) experiments \cite{zheng01,mito01}.
In both the compounds, anisotropic AFM spin fluctuations develop in the normal state.
The $T$ variation of $1/T_1$ in CeIrIn$_5$ pointed to the presence of anisotropic AFM  spin fluctuations near the QCP \cite{zheng01}.
Systematic study on the family of CeMIn$_5$ is promising in unraveling possible role of spin fluctuations for the emergence of superconductivity.

In this paper, we report results of In-NQR at zero field ($H$ = 0) and Co-NMR under a magnetic field $H = 1.1$ T on  single crystals of CeCoIn$_5$, focusing on the characteristics of spin fluctuations in the normal state.
In the normal state at $H = 0$, $1/T_1$ varies nearly as $T^{1/4}$ over a wide $T$ range.
This unique behavior is consistent with the anisotropic AFM spin-fluctuation's (SFs) model in case of the proximity to an AFM QCP\@.
The terminology of the anisotropic AFM spin fluctuations means here that a magnetic correlation length $\xi_{\rm plane}$ within the tetragonal plane develops more dominantly than $\xi_c$ does along the $c$-axis.
It is suggested that the value of $T_{\rm c} = 2.3$ K for CeCoIn$_5$ becomes larger than $T_{\rm c}$ = 0.4 K for CeIrIn$_5$ because of approaching the AFM QCP\@.
The AFM spin fluctuations in CeCoIn$_5$ are suppressed by the magnetic field, reinforcing that CeCoIn$_5$ is closely located to the QCP\@.

\section{Experimental}
High-quality single crystals of CeCoIn$_5$ and LaCoIn$_5$ were grown by In self-flux method as described elsewhere \cite{shishido02}.
For the NMR measurement in CeCoIn$_5$, we used several pieces of single crystals stacked with a shape of flat plate which undergo a SC transition at $T_{\rm c} = $ 2.3 K\@.
CeCoIn$_5$ has two inequivalent In sites per a unit cell.
The $^{115}$In-NQR $T_1$ measurement at $H = 0$ was performed at the In(1) site located on the top and the bottom faces of the tetragonal unit cell at the NQR frequency $\nu_Q \sim$ 8.16 MHz \cite{curro01,kohori01}.
The Co-NMR $T_1$ was measured under $H$ = 1.1 T\@.
The quadrupole frequency of $^{59}$Co for CeCoIn$_5$ is estimated as $\nu_Q\sim$ 0.23 MHz from the Co-NMR spectra, so it is experimentally difficult to detect Co-NQR signals at $H$ = 0.
Note that a principal axis of the electric field gradient is parallel to the tetragonal $c$-axis at the In(1) and Co sites.
In order to deduce the $4f$-derived contribution of magnetic fluctuations, the In-NQR and Co-NMR $T_1$ measurements have been made on LaCoIn$_5$ at $H = 0$ and 0.9 T, respectively.
These measurements were carried out by a conventional phase-coherent laboratory-built pulsed NQR/NMR spectrometer.

\section{Results and Discussion}
Figure 1 indicates the $T$ dependence of $1/T_1$ along with the data on LaCoIn$_5$ at $H = 0$ for the lowest quadrupole transition ($\pm 1/2\leftrightarrow\pm 3/2$).  
The $T$ dependence of $1/T_1$ is consistent with the result reported previously \cite{kohori01}.
In the SC state, $1/T_1$ decreases rapidly without any coherence peak just below $T_{\rm c} = 2.3$ K\@.
In contrast to the suggestion by Kohori {\it et al.\ }\cite{kohori01}, we remark that a simple behavior of $1/T_1\propto T^3$ is not valid, presumably affected by the close proximity to the AFM QCP\@.
Actually, a possibility has raised that the anomalous power-law $T$ behavior in London penetration depth at low $T$ well below $T_{\rm c}$ is due to an extension of quantum criticality into the SC state \cite{ozcan02}.
In this context, a deviation from  $1/T_1 \propto T^3$ allows us to address what happens in system located very close to AFM QCP when it becomes superconducting.
In order to clarify this, further experiments under pressure are now in progress.

\begin{figure}[t]
\begin{center}
\includegraphics[scale=0.42]{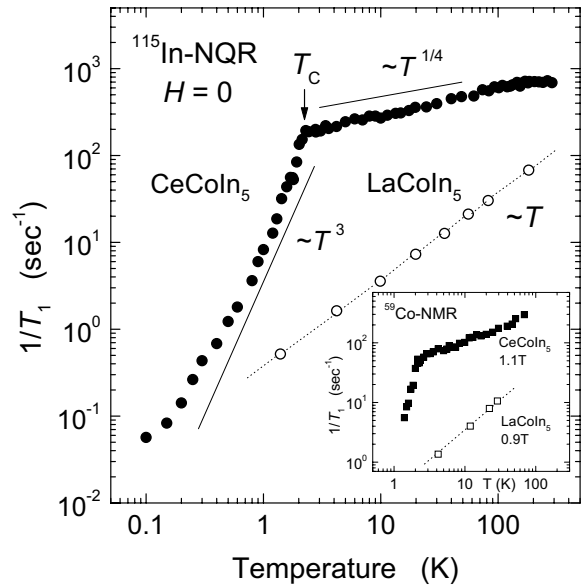}
\end{center}
\caption[]{$T$ dependence of $1/T_1$ of $^{115}$In in CeCoIn$_5$ along with  the data in LaCoIn$_5$ at $H = 0$.
The inset indicates the $T$ dependence of $1/T_1$ of $^{59}$Co at $H = 1.1$ T in CeCoIn$_5$ along with the data at $H =0.9$ T in LaCoIn$_5$\@.}
\end{figure}

Here, we focus on the characteristics of spin fluctuations in the normal state.
$1/T_1$ in LaCoIn$_5$ is proportional to $T$ as seen in Fig.~1\@.
In CeCoIn$_5$, however, $1/T_1$ is faster than that in LaCoIn$_5$\@.
It is remarkable that $(1/T_1)_{4f}$ varies nearly as $T^{1/4}$ in a wide $T$ range, where $(1/T_1)_{4f}$ corresponds to the $4f$ contribution of relaxation by subtracting a value of $1/T_1$ in LaCoIn$_5$ at each temperature.
This novel power-law behavior in $(1/T_1)_{4f}$ points to the presence of strong AFM spin fluctuations.
Above $T^* \sim$ 150 K, $1/T_1$ is almost independent of temperature, exhibiting that the system is in a localized regime.
It is known in HF systems that $T^*$ is scaled to the quasielastic linewidth in neutron-scattering spectrum, leading to a tentative estimation of the bandwidth of HF state.

Kohori {\it et al.\ }claimed that $1/T_1\sim T^{1/4}$-like dependence would arise from the spin fluctuations at the AFM QCP for isotropic three dimensional (3D) systems.
The SFs theory for strongly correlated electron systems near an AFM instability revealed that $1/T_1T\propto\sqrt{\chi_Q(T)}$ and $1/T_1T\propto\chi_Q(T)$ for 3D and 2D systems, respectively by assuming an expansion of dynamical susceptibility as $\chi(Q+q, \omega)^{-1}=\chi_Q^{-1}+a(q_x^2+q_y^2+q_z^2)-i\omega/\Gamma$.
Here, the staggered susceptibility at the AFM propagation vector $q = Q$, $\chi_Q(T)$, follows the Curie-Weiss law above the N\'eel temperature, namely, $\chi_Q(T) \propto 1/(T+\theta)$.
Note $\theta=0$ K at the AFM QCP\@.
At the AFM QCP or $\theta= 0$ K for 3D systems, $1/T_1$ is expected to be proportional to $\sqrt{T}$ at high $T$ and varies as $T^{1/4}$ for $t$ smaller than about $10^{-2}$ \cite{ishigaki}.
Here, $t$ is the reduced temperature defined by $t = T/T^*$.
It is hence unexpected for isotopic 3D systems to vary nearly as $1/T_1\sim T^{1/4}$ as observed in CeCoIn$_{5}$ over such the wide $T$ range.

We argue, therefore, this novel future in terms of the anisotropic AFM-SFs model due to the layered structure.
Let us now consider a case where a magnetic correlation length ($\xi_c$) along the $c$-axis for AFM spin fluctuations remains in a much shorter range than $\xi_{\rm plane}$ within the basal plane.
This is then modeled by assuming $\chi(Q+q)^{-1}=\chi_Q^{-1}+a_1(q_x^2+q_y^2)+a_2q_z^4$ instead of $\chi(Q+q)^{-1} = \chi_Q^{-1}+a(q_x^2+q_y^2+q_z^2)$.
In this model, $1/T_1T\propto\chi_Q(T)^{3/4} \propto 1/(T+\theta)^{3/4}$ is derived \cite{lacroix96}.
Recently, another calculation by assuming $\chi(Q+q)^{-1}=\chi_Q^{-1}+a(q_x^2+q_y^2+r_zq_z^2)$ with being $r_z$ an anisotropy parameter has shown that physical properties are interpolated smoothly between 2 and 3 dimensions \cite{kondo02}.
This result may be consistent with the claim that the observed $T$ dependence of $1/T_1\sim T^{1/4}$ in a wide $T$ range is due to anisotropic AFM SFs, since this $T$ variation is intermediate behavior between those for 2D and 3D systems.

\begin{figure}[t]
\begin{center}
\includegraphics[scale=0.44]{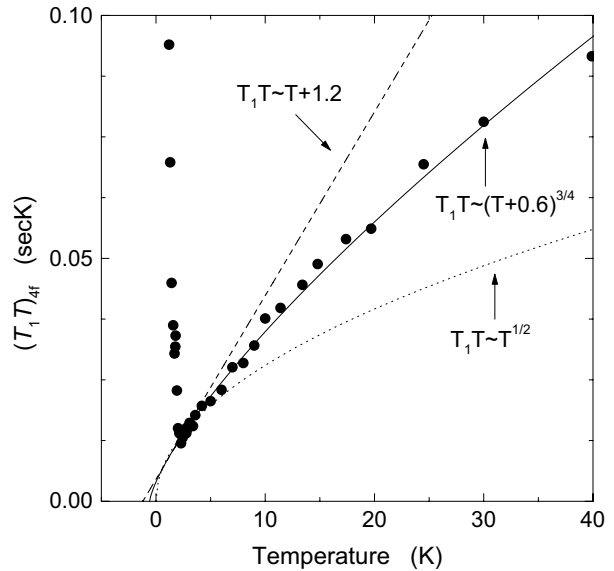}
\end{center}
\caption[]{$T$ dependence of $(T_1T)_{4f}$ at $H = 0$ in CeCoIn$_5$\@.
Here $(1/T_1T)_{4f} = (1/T_1T)_{\rm CeCoIn_5}-(1/T_1T)_{\rm LaCoIn_5}$.
The solid line indicates the relation $T_1T\propto (T+\theta)^{3/4}$ with $\theta=0.6$ K in terms of the anisotropic AFM spin-fluctuations (SFs) model (see the text).
The dotted and broken line correspond to the respective relations in terms of the isotropic 2D- and 3D-AFM-SFs models.}
\end{figure}

The $T$ dependence of $(T_1T)_{4f}$ is shown in Fig.~2\@.
The data are consistent with the anisotropic SFs model with $\theta= 0.6$ K as indicated by the solid line in the figure.
Neither the 3D nor 2D SFs models are applicable as indicated by the dotted or broken lines.
The value of $\theta = 0.6$ K close to zero demonstrates that CeCoIn$_5$ is closely located to the AFM QCP\@.
This anisotropic SFs model was already shown to interpret well the $T_1$ results of CeIrIn$_5$ with $\theta$ = 8 K that is significantly larger than that for CeCoIn$_5$ \cite{zheng01}.
Also the recent inelastic neutron experiment revealed that $\xi_{\rm plane}$ is larger than $\xi_c$ when approaching $T_{\rm N} = 3.5$ K in the HF antiferromagnet CeRhIn$_5$ at ambient pressure \cite{bao02}.
These magnetic correlations in CeMIn$_5$ are thus commonly characterized by the anisotropic AFM spin fluctuations.

We now discuss the relation between the value of $T_{\rm c}$ and what extent is closer to the AFM QCP in CeCoIn$_5$ and CeIrIn$_5$\@.
This closeness to the AFM QCP is evaluated from the value of $\theta$.
It is noteworthy that the value of $T_{\rm c}$ becomes larger as approaching the AFM QCP, since $T_{\rm c}$ increases as 0.4 K and 2.3 K as $\theta$ decreases as 8 K and 0.6 K for CeIrIn$_5$ and CeCoIn$_5$\@.
This suggests an intimate interrelation between the AFM spin fluctuations and the onset of $d$-wave superconductivity in these compounds.
Pagliuso {\it et al.\ }reported that the value of $T_{\rm c}$ goes up as the lattice parameter $c/a$ increases \cite{pagliuso}.
In this context, the increase in $c/a$ might enhance the 2D character of AFM spin fluctuations leading to the enhancement in $T_{\rm c}$\@.

Over a long decade, it is currently proposed that the value of $T_{\rm c}$ could be enhanced as the dimensionality in electronic structure is reduced from 3D to 2D regime \cite{monthoux,nakamura}.
Consistently with this suggestion, the pressure-induced SC transition temperature in CeIn$_3$, that forms in the 3D cubic structure, is one order smaller than $T_{\rm c} \sim$ 2.2 K in CeRhIn$_5$ that forms in the layered structure.
Note that, in CeIn$_3$, a $T_1T$ = const behavior is observed over a wide $T$ range above $T_{\rm c}$ in a pressure range where the AFM order is suppressed and the superconductivity appears \cite{shinji02, kohori02}.
It is, therefore, indicated that AFM spin fluctuations are absent in this pressure range in the normal state of CeIn$_3$\@.
By contrast, the AFM spin fluctuations in CeRhIn$_5$ continue to develop down to $T_{\rm c}$, exhibiting a character close to the AFM QCP \cite{mito01}.
The record high value of $T_{\rm c}$ in CeCoIn$_5$ at $P$ = 0 and in CeRhIn$_5$ under pressure to date is likely due to the existence of strong AFM spin fluctuations that develop in the vicinity of QCP\@.

Next, we move on to the magnetic field effect on the AFM spin fluctuations at the close vicinity of QCP in CeCoIn$_5$\@.
Since an energy scale is anticipated to be very reduced around a QCP, it is expected that the criticality at the AFM QCP is intimately influenced by applying small magnetic field.
As a matter of fact, YbRh$_2$Si$_2$ and CeRu$_2$Si$_2$, which are anticipated  to be located near a QCP, show the strong field dependence of SFs \cite{ishida98,ishida02,rossat-mignod}.
A recent theory also predicts that a strong magnetic field dependence should be observed at the QCP \cite{si}.

\begin{figure}[htbp]
\begin{center}
\includegraphics[scale=0.44]{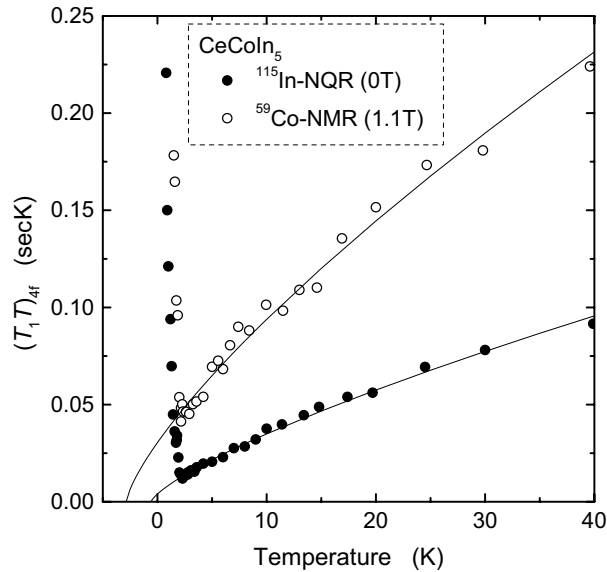}
\end{center}
\caption[]{$(T_1T)_{4f}$ vs $T$ plot at $H = 1.1$ T along with the $(T_1T)_{4f}$ vs $T$ plot at $H = 0$ T\@.
The solid lines indicate the relations $T_1T \propto (T+2.5)^{3/4}$ at $H =$ 1.1 T and $T_1T \propto (T+0.6)^{3/4}$ at $H$ = 0 T based on the anisotropic spin-fluctuation's model (see the text).
}
\end{figure}

In order to address possible $H$-induced change in AFM-SFs characteristics, we show the $T$ dependences of $1/T_1T$ in CeCoIn$_5$ and LaCoIn$_5$, in the inset of Fig.~1, that were measured at $H$ = 1.1 T parallel to the $c$-axis via the Co-NMR experiments.
Figure 3 indicates the $(T_1T)_{4f}$ vs $T$ plot at $H$ = 1.1 T along with the $(T_1T)_{4f}$ vs $T$ plot at $H$ = 0 T\@. 
The respective data at $H =$ 0 T and 1.1 T are consistent with the anisotropic SFs model with $\theta=$ 0.6 and 2.5 K as indicated by the solid lines.
Interestingly, the application of magnetic field makes the value of $\theta$ increase from 0.6 K to 2.5 K, making it away from the proximity to the AFM QCP\@.
This result reinforces that the CeCoIn$_5$ is closely located to the AFM QCP\@.

\section{Conclusion}

In conclusion, the measurements of the nuclear spin-lattice relaxation rate $1/T_1$ by means of the $^{115}$In-NQR and $^{59}$Co-NMR have revealed that the magnetic nature in CeCoIn$_5$ is characterized by the strong AFM spin fluctuations in the vicinity of QCP\@.
At zero magnetic field, the anomalous $T$ dependence of $(1/T_1)_{4f}$ is well explained by the anisotropic spin-fluctuations model that predicts the relation $T_1T \propto (T+\theta)^{3/4}$ with $\theta = 0.6$ K\@.
Noting that this model was applied to CeIrIn$_5$ with $\theta=8$ K, we have suggested that $T_{\rm c}$ rises in CeCoIn$_5$ because of approach to the AFM QCP where $\theta = 0$ K\@.
The application of magnetic field makes the value of $\theta$ increase from 0.6 K to 2.5 K, making it away from the QCP\@.
This result also suggests that CeCoIn$_5$ is closely located to the QCP\@.

\section*{Acknowledgements}

We would like to thank K.~Ishida for valuable comments and discussion.
This work was supported by the COE Research grant from MEXT of Japan (Grant No.\ 10CE2004).
One of the authors (Y.K.) was supported by the JSPS\@.

\end{document}